\documentclass[11pt]{article}

\usepackage{html,amsmath,amssymb,amsfonts,epsfig,cancel,cite,longtable,caption,siunitx,array,color,booktabs,setspace}

\usepackage{dsfont}

\newlength{\xtrawidth}
\newlength{\xtraheight}
\renewcommand\baselinestretch{1.25}

\numberwithin{equation}{section}

\newcommand{\be}{\begin{equation}}
\newcommand{\ee}{\end{equation}}
\newcommand{\beq}{\begin{equation}}
\newcommand{\eeq}{\end{equation}}
\newcommand{\ba}{\begin{array}}
\newcommand{\ea}{\end{array}}
\newcommand{\bea}{\begin{eqnarray}}
\newcommand{\eea}{\end{eqnarray}}
\newcommand{\bean}{\begin{eqnarray*}}
\newcommand{\eean}{\end{eqnarray*}}

\newcommand{\fref}[1]{Figure~\ref{#1}}

\newcommand{\IC}{\mathbb{C}}
\newcommand{\IP}{\mathbb{P}}

\newcommand{\IH}{\mathbb{H}}

\newcommand{\IZ}{\mathbb{Z}}

\newcommand{\cN}{{\cal N}}

\newcommand{\cA}{{\cal A}}
\newcommand{\cB}{{\cal B}}
\newcommand{\cC}{{\cal C}}

\newcommand{\cV}{{\cal V}}

\newcommand{\fn}{\footnotesize}

\def\cjn1{{\cA, \cC^*\otimes \wedge^j \cN^*}}
\def\bjn1{{\cA, \cB^*\otimes \wedge^j \cN^*}}
\def\vjn1{{\cA, \cV^*\otimes \wedge^j \cN^*}}
\def\cjn2{{\cA, \cC\otimes \wedge^j \cN^*}}
\def\bjn2{{\cA, \cB\otimes \wedge^j \cN^*}}
\def\vjn2{{\cA, \cV\otimes \wedge^j \cN^*}}

\newcommand{\cicy}[2]{\begin{matrix} #1\end{matrix}\!\left[\begin{matrix}#2 \end{matrix}\right]}

\newcommand{\capt}[3]{\parbox{#1}{\renewcommand{\baselinestretch}{1.0}
                                                           \caption{\label{#2}\small\it #3}}}

\definecolor{db}{rgb}{0,0,0.8}
\definecolor{dg}{rgb}{.1,0.5,.5}
\definecolor{dr}{rgb}{0.9,0,0}

\begin{document}

\title{{\LARGE \bf The Family Problem: \\[1mm] Hints from Heterotic Line Bundle Models\\[4mm]}}

\author{
Andrei Constantin$^{1,2}$,
Andre Lukas$^2$, 
Challenger Mishra$^2$
}
\date{}
\maketitle
\thispagestyle{empty}
\begin{center} { 
${}^1\,$\itshape Department of Physics and Astronomy, Uppsala University, \\ 
       SE-751 20, Uppsala, Sweden\\[0.3cm]
${}^2\,$Rudolf Peierls Centre for Theoretical Physics, Oxford University,\\
       1 Keble Road, Oxford, OX1 3NP, U.K.}\\
       
\end{center}

\vspace{1cm}
\abstract
\noindent 
Within the class of heterotic line bundle models, we argue that $\mathcal N=1$ vacua which lead to a small number of low-energy chiral families are preferred. By imposing an upper limit on the volume of the internal manifold, as required in order to obtain finite values of the four-dimensional gauge couplings, and validity of the supergravity approximation we show that, for a given manifold, only a finite number of line bundle sums are consistent with supersymmetry. By explicitly scanning over this finite set of line bundle models on certain manifolds we show that, for a sufficiently small volume of the internal manifold, the family number distribution peaks at small values, consistent with three chiral families. The relation between the maximal number of families and the gauge coupling is discussed, which hints towards a possible explanation of the family problem.
\vskip 3cm
{\hbox to 7cm{\hrulefill}}
\noindent
{\fn andrei.constantin@physics.uu.se\\ lukas@physics.ox.ac.uk\\ challenger.mishra@physics.ox.ac.uk} 
\newpage


%
%
\section{Introduction}
String theory provides a natural explanation for the family replication observed in nature: Kaluza-Klein compactification from ten to four dimensions generically leads to low-energy multiplets appearing with a certain multiplicity which is governed by a topological number associated to the compactification \cite{Witten:1983ux, Candelas:1985en}. Despite the mathematical beauty of this mechanism and the qualitative solution of the family problem it provides, there is no immediate prediction for the number of families. Different compactifications lead to different family numbers and, although three families can be achieved by appropriate model building choices, a wide range of values can be obtained.

In this paper, we would like to study the question of family number in string theory in the context of heterotic line bundle models, a class of models introduced in Refs.~\cite{Anderson:2011ns,Anderson:2012yf,Anderson:2013xka} and further developed in Refs.~\cite{He:2013ofa, Buchbinder:2013dna, Buchbinder:2014qda, Buchbinder:2014sya, Anderson:2014hia, Buchbinder:2014qca, Nibbelink:2015ixa, Nibbelink:2015vha}, see also the earlier studies \cite{Distler:1987ee, Blumenhagen:2005ga}. These models are based on Calabi-Yau compactifications of the heterotic string and vector bundles with Abelian structure group, that is, line bundle sums. As has been shown in Refs.~\cite{Anderson:2011ns,Anderson:2012yf,Anderson:2013xka}, these models are very well under control from a model building point of view and large numbers of quasi-realistic models with a standard model spectrum can be obtained. For these reasons, heterotic line bundle models provide a useful setting to study the family number problem. One observation, made in Ref.~\cite{Anderson:2013xka}, is of particular importance for this discussion. Based on extensive computer scans, it was noted that, for a given Calabi-Yau threefold $X$, the class of Abelian bundles which correspond to $\cN=1$ supersymmetric vacua in the interior of the K\"ahler cone and lead to three chiral families is finite.

In Ref.~\cite{Buchbinder:2013dna}, this finiteness result was shown analytically, based on two assumptions on the K\"ahler moduli of the theory: the K\"ahler moduli have to be sufficiently away from the boundary of the K\"ahler cone and the Calabi-Yau volume is finite~\footnote{An infinite class of heterotic line bundle models has been constructed in a recent paper~\cite{Nibbelink:2015ixa}, by including the boundary of the K\"ahler cone. In the present paper, the K\"ahler moduli are always constrained to the interior of the K\"ahler cone, so we do not encounter such infinite families of models.}. The first of these assumptions stems from the requirement that supergravity remains valid at the chosen locus in moduli space and the second one is motivated by the finiteness of the four-dimensional couplings constants. Interestingly, the proof of this statement does not require imposing a fixed number of families - supersymmetry and anomaly cancellation together with the aforementioned constraints on the K\"ahler moduli space are sufficient. This suggests that the class of bundles that lead to consistent $\cN=1$ heterotic vacua with an arbitrary number of chiral families is also finite, in line with the conjecture made in Ref.~\cite{Douglas:2006jp}. 

Provided this finiteness result holds, it is clear there exists an upper bound  on the number of families for a given Calabi-Yau manifold. What is more, this upper bound  depends on and monotonically increases with the Calabi-Yau volume. Roughly, the more stringent the bound on the Calabi-Yau volume the fewer line bundle models can be supersymmetric in the so-prescribed portion of K\"ahler moduli space and the lower the bound on the number of families. 
This relation between the maximal number of families and the Calabi-Yau volume is rather surprising and hints at a possible explanation of the family problem: The number of families is small because the four-dimensional coupling constants have sizeable, finite values which require a relatively small Calabi-Yau volume.

The aim of the present paper is to establish this connection between the Calabi-Yau volume and the number of families for heterotic line bundle models in detail. We begin by reviewing the basic model-building set-up in Section~\ref{sec:setup} and by discussing coupling constants in Section~\ref{sec:pheno}. In Section~\ref{sec:analytic}, we revisit the proof presented in Ref.~\cite{Buchbinder:2013dna} and derive a semi-analytical formula for the maximal number of families. Section~\ref{sec:scan} presents an explicit construction of heterotic line bundle models with  low-energy gauge group $SU(5)$ and a variable number of families and determines the maximal number of generations as a function of the Calabi-Yau volume. We conclude in Section~\ref{sec:conclusion}.

\section{Model building setup} \label{sec:setup}
Heterotic Calabi-Yau models with $\mathcal N=1$ supersymmetry are specified by a Calabi-Yau three-fold, $X$, and a vector bundle $V\rightarrow X$ with structure group contained in $E_8\times E_8$. For a consistent vacuum the bundle $V$ needs to satisfy two further conditions: it needs to obey the heterotic anomaly cancellation condition and, in order to preserve $\mathcal N=1$ supersymmetry, it needs to be poly-stable with slope zero. Since both conditions are crucial for the subsequent discussion we would now like to briefly review them in turn.\\[3mm]
To discuss poly-stability we introduce the slope 
\beq
\mu({\cal F}) = \frac{c_1({\cal F})\cdot J^2 }{{\rm rk}({\cal F})}~.
\eeq
of a coherent sheaf ${\cal F}$, where $J$ is the K\"ahler form of $X$. A bundle $V$ with a simple structure group is called slope-stable iff $\mu({\cal F})<\mu(V)$ for all coherent sheafs ${\cal F}\subset V$ with $0<{\rm rk}({\cal F})<{\rm rk}(V)$.  Further, a direct sum bundle $V=V_1\oplus\cdots\oplus V_n$ is called poly-stable if all summands $V_i$ are slope-stable and if they have the same slope, that is, $\mu(V_1)=\cdots=\mu(V_n)=\mu(V)$. In the present context we are interested in rank $n$ line bundle sums
\beq\label{eq:V}
V=\bigoplus_{a=1}^{n} L_a\quad\mbox{satisfying}\quad c_1(V)=\sum_{a=1}^nc_1(L_a)\stackrel{!}{=}0\; ,
\eeq
which have a typical structure group $S(U(1)^n)$. 
Specifically, we will focus on rank five bundles, so $n=5$. In this case, the structure group can be embedded into $E_8$ via the subgroup chain $S(U(1)^5)\subset SU(5)\subset E_8$ which leads to a low-energy GUT group $SU(5)\times S(U(1)^5)$. Quasi-realistic standard models can be obtained from these GUT models after dividing by a freely-acting discrete symmetry (in cases when $X$ is simply-connected) and including a suitable Wilson line. 

Due to the constraint on the rank of the sub-sheaf ${\cal F}$, line bundles are automatically slope-stable. All we have to require for a supersymmetric line bundle sum is, therefore, the vanishing of all slopes, that is
\begin{equation}
 \mu(L_a)=c_1(L_a)\cdot J^2\stackrel{!}{=}0 \label{slopes0}
\end{equation} 
for all $a=1,\ldots ,n$. Note that, for fixed line bundles $L_a$, these are conditions on the K\"ahler form, $J$, of the Calabi-Yau manifold $X$. In practice, we have to check if the slopes of all line bundles can simultaneously vanish somewhere in (the interior of) the K\"ahler cone of $X$.

A slope poly-stable bundle $V$ automatically satisfies a positivity condition on the second Chern class \cite{huybrechts2010geometry} which is given by
\beq\label{eq:c2Bogomolov}
\int_X c_2(V)\wedge J \geq 0\; ,
\eeq
and is known as the Bogomolov bound. Here, $J$ is any K\"ahler form for which $V$ is poly-stable. For a line bundle sum~\eqref{eq:V} the second Chern class is given by
\begin{equation}
 c_2(V)=-\frac{1}{2}\sum_{a=1}^nc_1(L_a)^2\; . \label{c2c1}
\end{equation}\\[3mm] 
In order to be able to satisfy the anomaly cancellation condition we require that
\begin{equation}
 c_2(TX)-c_2(V)\in \mbox{Mori cone of }X\; . \label{inMori}
\end{equation} 
Provided this condition is satisfied, we can always saturate the anomaly condition by adding a suitable hidden sector of five-branes (or a suitable bundle in the other, hidden $E_8$ factor or a combination of five branes and hidden bundle). In practice, it will be useful to introduce an integral basis $C_i$, where $i=1,\ldots ,h^{1,1}(X)$, of holomorphic curves for the second homology of $X$ and a corresponding dual basis, $J_i$, of $H^{1,1}(X)$. Then, the K\"ahler from can be expanded as
\begin{equation}
 J=\sum_{i=1}^{h^{1,1}(X)}t^iJ_i\; ,
\end{equation}
where $t^i$ are the K\"ahler moduli. For our examples, the K\"ahler cone of $X$ is simply characterised by all $t^i>0$ and the holomorphic curves $C_i$ generate the Mori cone. The latter property means that the anomaly condition~\eqref{inMori} can be re-written as
\begin{equation}
 \left(c_2(TX)-c_2(V)\right)\cdot J_i\geq 0\; ,\label{anomcond}
\end{equation}
for all $i=1,\ldots  ,h^{1,1}(X)$.\\[3mm] 
Another crucial quantity for our discussion is the number of chiral families, given by the index
\beq
N_{\rm gen} = -\text{ind}(V) = \frac{1}{2}\int_X c_3(V)~, \label{Ngen}
\eeq
where we have used $c_1(V) =0$ in the last step.

Equations \eqref{eq:V}, \eqref{eq:c2Bogomolov} and \eqref{anomcond} constrain the first and the second Chern classes of~$V$, while Eq.~\eqref{slopes0} guarantees the existence of points in the K\"ahler moduli space of $X$ for which~$V$ is poly-stable with zero slope. In the following we would like to provide evidence that the class of line bundle models on a given Calabi-Yau manifold, subject to these conditions on the first and second Chern classes is finite. Of course, this implies the existence of an upper bound on $N_{\rm gen}$ for a given Calabi-Yau manifold. We will prove finiteness of the class after excluding a finite neighbourhood at the boundary of the K\"ahler moduli space and requiring the Calabi-Yau volume to be finite. These restrictions are motivated by the validity of the supergravity approximation and the finiteness of the four-dimensional  couplings, as mentioned earlier. On the other hand, the results of our automated scans indicate that the class remains finite even without these restrictions on the K\"ahler moduli space. 

\section{K\"ahler moduli space and low-energy coupling constants}\label{sec:pheno}
For a supersymmetric line bundle sum, the slope zero conditions~\eqref{slopes0} must have a common solution in the K\"ahler cone of the Calabi-Yau manifold $X$. However, from a physical point of view, the acceptable locus in the K\"ahler cone is further restricted by the values of low-energy coupling constants and the requirement that the supergravity approximation be consistent. We would like to discuss the interplay between those physical restrictions in K\"ahler moduli space and the slope zero conditions. Unification of gauge couplings, including the gravitational coupling, in the heterotic string is most naturally realised in the strong-coupling limit~\cite{Witten:1996mz}, described by 11-dimensional Horava-Witten theory~\cite{Horava:1996ma}. For this reason, we will be working in the 11-dimensional theory and measure all internal volumes using the relevant part of the 11-dimensional metric. 

The 11-dimensional Newton constant $\kappa_{11}$ and the 11-dimensional Planck length $l$ are related by $4\pi\kappa_{11}=(2\pi l)^9$ and we also introduce the six-dimensional coordinate volume $v=(2\pi l)^6$. The dimensionless K\"ahler moduli $t^i$ and the triple intersection numbers are defined by
\begin{equation}
 t^i=\frac{1}{(2\pi l)^2}\int_{C_i}J\; ,\qquad d_{ijk}=\frac{1}{v}\int_XJ_i\wedge J_j\wedge J_k\; ,
\end{equation}
where $J$ is the K\"ahler form associated to the 11-dimensional metric.  Hence, the K\"ahler moduli $t^i$ measure the volume of the holomorphic cycles $C_i$ in units of the 11-dimensional Planck length. As usual, we introduce the pre-potential 
\begin{equation}
\kappa=6{\cal V}=d_{ijk}t^it^jt^k\; ,
\end{equation}
where ${\cal V}$ is the volume in units of the coordinate volume $v$, so that the physical volume is given by
\begin{equation}
 V_{\rm phys}=\frac{1}{3!}\int_XJ\wedge J\wedge J=v{\cal V}\; .
\end{equation}   
The four-dimensional GUT coupling constant can then be written as~\cite{Witten:1996mz}
\begin{equation}
 \alpha_{\rm GUT}=\frac{(2\pi l)^6}{2V_{\rm phys}}=\frac{1}{2{\cal V}}\; .
\end{equation} 

We are now ready to discuss the relevant physical restrictions on the K\"ahler moduli space. Validity of the supergravity approximation requires the volume of all cycles $C_i$ to be larger than one in 11-dimensional Planck units which implies restricting the K\"ahler moduli as
\beq\label{eq:t>1}
t^i\stackrel{!}{>}1\quad \text{for all}\quad i=1,\ldots,h^{1,1}(X)\; .
\eeq

Further, in order to match the GUT coupling constant we should require that the value
\begin{equation}
 {\cal V}=\frac{1}{6}d_{ijk}\,t^i\,t^j\,t^k\stackrel{!}{\simeq}\frac{1}{2\alpha_{\rm GUT}}\simeq 12 \label{unival}
\end{equation} 
can be realised in the restricted K\"ahler moduli space defined by Eqs.~\eqref{eq:t>1} intersected with the locus obtained by imposing the slope-zero conditions~\eqref{slopes0}. 
If this intersection region is non-empty, it is in fact unbounded, since the slope-zero conditions~\eqref{slopes0} are homogeneous in the K\"ahler moduli $t^i$ and are, therefore, satisfied on rays (indicated by the red lines in Fig.~\ref{fig:Kahler}) in K\"ahler moduli space. Accordingly, the Calabi-Yau volume assumes arbitrarily large values over this region, while being bounded from below. Thus, if the minimal value attained by the Calabi-Yau volume is larger than the unification value, the model is ruled out. 

In practice, this means that we can search for models with realistic gauge couplings by restricting the K\"ahler moduli such that ${\cal V}\leq {\cal V}_{\rm max}$. From Eq.~\eqref{unival}, the physically relevant value of $V_{\rm max}$ should be approximately $12$. However, later on we will allow a wider range for $V_{\rm  max}$ values in order to study the dependence of the maximal number of generations on the value of the gauge coupling. The situation in K\"ahler moduli space for two K\"ahler moduli is schematically indicated in Fig.~\ref{fig:Kahler}.
\begin{figure}[h]
\begin{center}
\includegraphics[width=5.9cm]{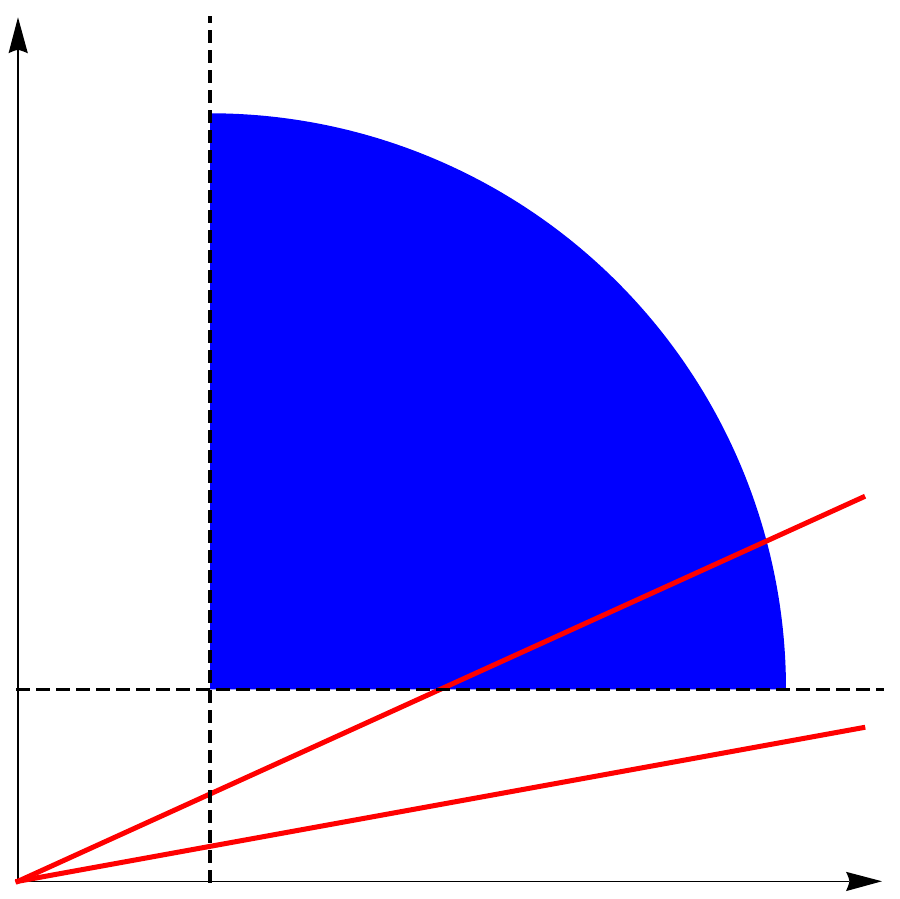}
\vskip 0.5cm
\capt{14cm}{fig:Kahler}{Sketch of the allowed region in K\"ahler moduli space. The slope-zero equations are invariant under a change $t^i\rightarrow a\, t^i$, hence a line bundle sum which is consistent at some point, is consistent along the entire ray containing that point. }
\end{center}
\end{figure}
\vskip -5mm\noindent
The K\"ahler cone corresponds to the positive quadrant and the blue region indicates the physically allowed region, which is detached from the boundaries of the K\"ahler cone due to the Eqs.~\eqref{eq:t>1} and is bounded~\footnote{A finite volume may not necessarily bound the K\"ahler moduli space if some intersection numbers are negative. Here we assume that the region in K\"ahler moduli space is indeed bounded by the finite volume requirement. This is certainly the case if all intersection numbers are positive as will be the case for our examples.} due to the condition $V\leq V_{\rm max}$. The rays defined by the slope-zero conditions~\eqref{slopes0} are indicated by the red lines in Fig.~\ref{fig:Kahler}. A given line bundle model is supersymmetric and consistent with the physical restrictions on the K\"ahler moduli space if the corresponding ray intersects the blue region in Fig.~\ref{fig:Kahler}.  In the next section we will show that this class of supersymmetric, physical line bundle sums is finite. 

\section{A semi-analytic bound}\label{sec:analytic}
We write the line bundles in Eq.~\eqref{eq:V} as $L_a={\cal O}_X({\bf k}_a)$, so that their first Chern class is given by $c_1(L_a)=k^i_aJ_i$. The integers $k_a^i$ are constrained by the conditions
\beq\label{eq:c1constraint}
\sum_{a=1}^{n} k^i_a = 0
\eeq
for all $i=1,\ldots,h^{1,1}(X)$ which are equivalent to $c_1(V)=0$. Further, from Eq.~\eqref{c2c1}, the second Chern class is given by
\beq
c_{2i}(V) = -\frac{1}{2}\sum_{a=1}^{n} d_{ijk}\,k^i_a\,k^j_a\;,
\eeq
and in order to be able to satisfy the anomaly cancellation condition we require that
\beq
 c_{2i}(V)\leq c_{2i}(TX)\;, \label{acons}
\eeq
in accordance with Eq.~\eqref{anomcond}. The slope zero conditions~\eqref{slopes0} take the form
\beq\label{eq:slopezero}
d_{ijk}\, k^i_a\, t^j\, t^k = 0\; ,
\eeq
for $i=1,\ldots,h^{1,1}(X)$, and these equations have to be simultaneously satisfied in the interior of the K\"ahler cone (here taken to be characterised by $t^i>0$ for all $i$). The question we would like to address is whether line bundle sums $V$ on a given Calabi-Yau manifold $X$, subject to the $c_1(V)=0$ constraint~\eqref{eq:c1constraint}, the anomaly constraint~\eqref{acons} and the slope zero conditions~\eqref{eq:slopezero} constitute a finite class. If they do, the number of generations 
\beq
 N_{\rm gen}=-{\rm ind}(V) = -\frac{1}{6}\,d_{ijk}\sum_{a=1}^{n} k^i_a\,k^j_a\,k^k_a\; .
 \eeq
for this class will also be finite and we are more specifically interested in any bounds on this number. The automated scans described in the next section indicate that the answer is in the affirmative, although it seems difficult to provide a general proof and derive a precise expression for the bound on $N_{\rm gen}$. 

However, we would like to provide an analytical finiteness argument under the assumption that the slope conditions~\eqref{eq:slopezero} are satisfied in the {\em physical} part of K\"ahler moduli space (as discussed in the previous section). First, we recall that the K\"ahler moduli space is equipped with a positive-definite metric \cite{Candelas:1990pi} 
\begin{equation}
G_{ij} = \frac{1}{2v{\cal V}} \int_X J_i\wedge \star J_j = -3 \left( \frac{\kappa_{ij}}{\kappa} - \frac{2\kappa_i\kappa_j}{3\kappa^2}\right) \; ,
\end{equation}
where $\kappa_i = d_{ijk}\,t^j\,t^k$ and $\kappa_{ij}=d_{ijk}\,t^k$. Due to the slope zero conditions~\eqref{eq:slopezero}, which can also be written as $\kappa_ik_a^i=0$, we obtain 
\begin{equation}\label{eq:bound1}
0< \sum_a {\bf k}_a^T G\, {\bf k}_a = -\frac{3}{\kappa} d_{ijk} \sum_a k_a^i\, k_a^j\, t^k = \frac{6}{\kappa}\, t^i\, {c}_{2i}(V) \leq \frac{6}{\kappa}  t^i\, {c}_{2i}(TX)  \leq \frac{6}{\kappa}|{\bf t}||c_{2}(TX)|\; .
\end{equation}
Introducing the modified moduli space metric $\widetilde G  = \kappa\, G /( 6 |{\bf t}|)$ (which is homogeneous of degree zero in $t^i$) this means that
\begin{equation}\label{kbound}
 \sum_a {\bf k}_a^T \widetilde{G}\, {\bf k}_a\leq |c_{2i}(TX)|\; . 
 \vspace{-9pt}
\end{equation} 
Since $\widetilde G$ is positive definite in the (interior of the) K\"ahler cone, Eq.~\eqref{kbound} seems to imply the existence of a bound on $|{\bf k}_a|$, and hence a bound on $N_{\rm gen}$. However, the K\"ahler metric can degenerate on the boundary of the K\"ahler cone which means that, in the interior of the K\"ahler cone, the eigenvalues of $\widetilde G$ cannot be bounded from below by a strictly positive number. Hence, if we allow solutions in the entire interior of the K\"ahler cone, Eq.~\eqref{kbound} does not provide an argument for finiteness. 

The situation improves when we strengthen our assumptions and demand a solution in the {\em physical} region of K\"ahler moduli space, as defined in the previous section and schematically indicated in Fig.~\ref{fig:Kahler}. In this case, the eigenvalues of $\widetilde{G}$ are bounded from below by the minimal (but strictly positive) eigenvalue $\lambda_{\rm min}>0$ of $\widetilde{G}$ over this physical region and the line bundle integers are bounded by
\begin{equation}
 \sum_a |{\bf k}_a|^2\leq\frac{|c_2(TX)|}{\lambda_{\rm min}}\; . \label{kbound1}
\end{equation} 
Note that the value of $\lambda_{\rm min}$ depends on the topology of the Calabi-Yau manifold (notable on the triple-intersection numbers) and the maximal volume ${\cal V}_{\rm max}$ used to define the physical region. For a given Calabi-Yau manifold an upper bound on the volume, $\lambda_{\rm min}$ can be determined, although we do not have an explicit analytic formula. Given Eq.~\eqref{kbound1} the line bundle entries are bounded by a value $|k_{\rm max}|$ which satisfies
\beq
|k_{\rm max}|~\leq~ \sqrt{ \frac{1}{2}\left( \frac{|c_2(TX)|}{\lambda_{\rm min}} - (h^{1,1}(X)-1)\right)}\; ,
\eeq
where we have taken into account the constraint $c_1(V)=0$ and we have assumed that the $k_a^i$ do not vanish for all $a$ and a given $i$. Since $t^i>1$, we have $ k^i_a  < | k_{\rm max} | \,t^i$ and
\beq
\sum_a d_{ijk}\,(k^i_a+| k_{\rm max} | t^i)\,(k^j_a+| k_{\rm max} |t^j)\,(k^k_a+| k_{\rm max} |t^k) > 0
\eeq
which, given the slope zero conditions~\eqref{eq:slopezero}, becomes 
\beq
N_{\rm gen} <  |k_{\rm max}|^3\, {\cal V}- |k_{\rm max}|c_{2i}(V)t^i\; .
\eeq
Given a finite volume ${\cal V}<{\cal V}_{\rm max}$, strict positivity of $\lambda_{\rm min}$ over the physical part of K\"ahler moduli space and the positivity condition \eqref{eq:c2Bogomolov}, this does indeed provide an upper bound for the number of generations on a given Calabi-Yau manifold.

\section{Computer scan results}\label{sec:scan}
In this section we present the results of an automated scan performed on several different Calabi-Yau three-folds for $SU(5)-$models with a variable number of generations.\footnote{Frequency distributions of models with various numbers of generations have been performed within other string contexts, such as orientifolds of Gepner models \cite{Dijkstra:2004cc, GatoRivera:2008zn}, intersecting D-brane models \cite{Gmeiner:2005vz}, heterotic constructions with free-fermions \cite{Faraggi:2006bc, Assel:2010wj}, heterotic Calabi-Yau compactifications with monad bundles \cite{Anderson:2008uw} and heterotic orbifold constructions \cite{Nilles:2014owa}} As we will see, these results provide evidence for the finiteness of line bundle models with vanishing slope in the interior of the K\"ahler cone and lead to an upper bound for the number of generations. We impose the constraints  \eqref{eq:V}, \eqref{slopes0}, \eqref{eq:c2Bogomolov} and \eqref{anomcond} and, in addition, we will require that, for each pair of indices $a<b$, ${\rm ind}(L_a\otimes L_b)\leq 0$, a condition necessary in order to project out all Higgs triplets after the inclusion of a Wilson line. 

Moreover, it would be both interesting and important to use these results in order to find all line bundle models with vanishing slope in the physical region of K\"ahler moduli space, for varying Calabi-Yau volume ${\cal V}$, and to determine an upper bound for the number of generations as a function of ${\cal V}$. However, in general this is not an easy task, as explained below. 

Finding the locus where the line bundle sum is poly-stable corresponds to simultaneously solving the quadratic equations \eqref{eq:slopezero} in the $t^i$ variables. This makes the process described above difficult to implement. However, things greatly simplify in the case in which the locus defined by the slope zero conditions is a ray, which corresponds to having the number of linearly independent $\mathbf{k}_a$ vectors equal to $h^{1,1}(X)-1$, i.e.,
\beq
{\rm rank}(k^i_a) = h^{1,1}(X)-1\label{rankconstraint}
\eeq

 In this case, after finding an arbitrary non-trivial solution to the slope zero equations (e.g.~using the routine \texttt{FindInstance} in Mathematica), the minimal value assumed by the Calabi-Yau volume in the region of interest in obtained by rescaling the solution such that all the $t_i$ parameters are greater than or equal to~1. 

For the tetra-quadric manifold discussed below most of the line bundle models are such that the poly-stability locus is a single ray. Excluding the small number of models which do not satisfy the condition~\eqref{rankconstraint} does not influence the qualitative conclusions of our discussion.

\subsection{The tetra-quadric manifold}
Tetra-quadric manifolds are simply connected hypersurfaces in a product of four $\IC\IP^1$ spaces, defined as the zero locus of a homogeneous polynomial that is quadratic in the homogeneous coordinates of each $\IC\IP^1$ space. Manifolds in this class have Euler number $\eta = - 128$ and Hodge numbers $h^{1,1}(X)=4$ and $h^{2,1}(X)=68$. This information is summarised by the following configuration matrix:
\begin{equation}
X~=~~
\cicy{\IC\IP^1 \\   \IC\IP^1\\ \IC\IP^1\\ \IC\IP^1}
{ ~2 \!\!\!\!\\
  ~2\!\!\!\! & \\
  ~2\!\!\!\! & \\
  ~2\!\!\!\!}_{-128}^{4,68}\
\end{equation}
The second cohomology is spanned by the four K\"ahler forms, $J_i$ of the $\IC\IP^1$ factors, pulled-back to the tetra-quadric, and we also introduce a dual basis, $\nu^i$ of the fourth cohomology. The triple intersection numbers have the following simple form
\beq
d_{ijk} = \int_X J_i\wedge J_j\wedge J_k = \begin{cases} 2 & \mbox{ if } i\neq j, j\neq k \\ 0 &\mbox{ otherwise } \end{cases}\; \label{tqisec}
\eeq
and they lead to the pre-potential
\begin{equation}
\kappa  = 6{\cal V}=12\left( t_1t_2t_3+t_1t_2t_4+t_1t_3t_4+t_2t_3t_4\right)\; .
\end{equation}
The K\"ahler cone is characterised by all $t^i>0$ while the Mori cone corresponds to positive linear combinations of the $\nu^i$. The second Chern class of the tangent bundle of the tetra-quadric, in the basis $\nu^i$, is given by
\begin{equation}
 c_2(TX)=(24,24,24,24)\; . \label{tqc2}
\end{equation} 
The tetraquadric manifold admits smooth quotients by finite groups $\Gamma$ of orders $|\Gamma|=2, 4, 8$ and $16$. Specifically, these groups are $\Gamma= \IZ_2,\,\IZ_2\times \IZ_2,\, \IZ_4$, $\IZ_2\times \IZ_4,\,\IZ_8,\,\IH,\,\IZ_4\times \IZ_4,\,\IZ_4 \rtimes \IZ_4,\,\IZ_8\times \IZ_2,\,\IZ_8\rtimes \IZ_2,\,\IH\times \IZ_2$ \cite{Braun:2010vc}. The physical model with standard model group and three chiral families is defined on the quotient manifold $X/\Gamma$ (provided the bundle $V\rightarrow X$ descends to the quotient) while, in practice, we calculate the chiral asymmetry for the upstairs model. The two chiral asymmetries of upstairs and downstairs model are related by 
\begin{equation}
N_{\text{gen}}(X/\Gamma) = \frac{N_{\text{gen}}(X)}{|\Gamma|}\stackrel{!}{=}3\; .
\end{equation}
Hence, three chiral families in the downstairs model require
\beq
N_{\rm gen}(X)=6,12,24,48 \label{tqNgen}
\eeq
families in the upstairs model, depending on which symmetry $\Gamma$ is involved. Also, the value imposed on the volume from Eq.~\eqref{unival} should be considered in the downstairs model, hence for the upstairs model we require ${\cal V} \simeq 12\, |\Gamma|$.

We have scanned over a large number ($\sim 10^{35}$) of rank five line bundle sums and have extracted all models which satisfy the anomaly condition~\eqref{acons} and the slope zero conditions~\eqref{eq:slopezero} in the interior of the K\"ahler cone. This has been done for increasing sizes of the line bundle integers $|k_a^i|$, until no further models could be found. In practice, this means that we have found all consistent rank five line bundle models on the tretra-quadric. Within this set, the generation number does not exceed the value $N_{\text{gen}} = 126$ and the distribution of the generation number  is shown in~\fref{fig:plot7862}. We can further focus on the sub-sets of these models which satisfy the slope zero condition in the physical region of K\"ahler moduli space, corresponding to a certain maximal Calabi-Yau volume ${\cal V}_{\rm max}$. The distribution of the generation number for these sub-sets is indicated by the colour-coding in~\fref{fig:plot7862}. In particular, for the unification value ${\cal V}\simeq 12\,|\Gamma|$ from Eq.~\eqref{unival}, it can be seen from \fref{fig:plot7862v2} that $N_{\text{gen}}\lesssim 40$ which is well in line with the required upstairs values in Eq.~\eqref{tqNgen}.
\begin{figure}[!h]
\begin{center}
\includegraphics[width=12cm]{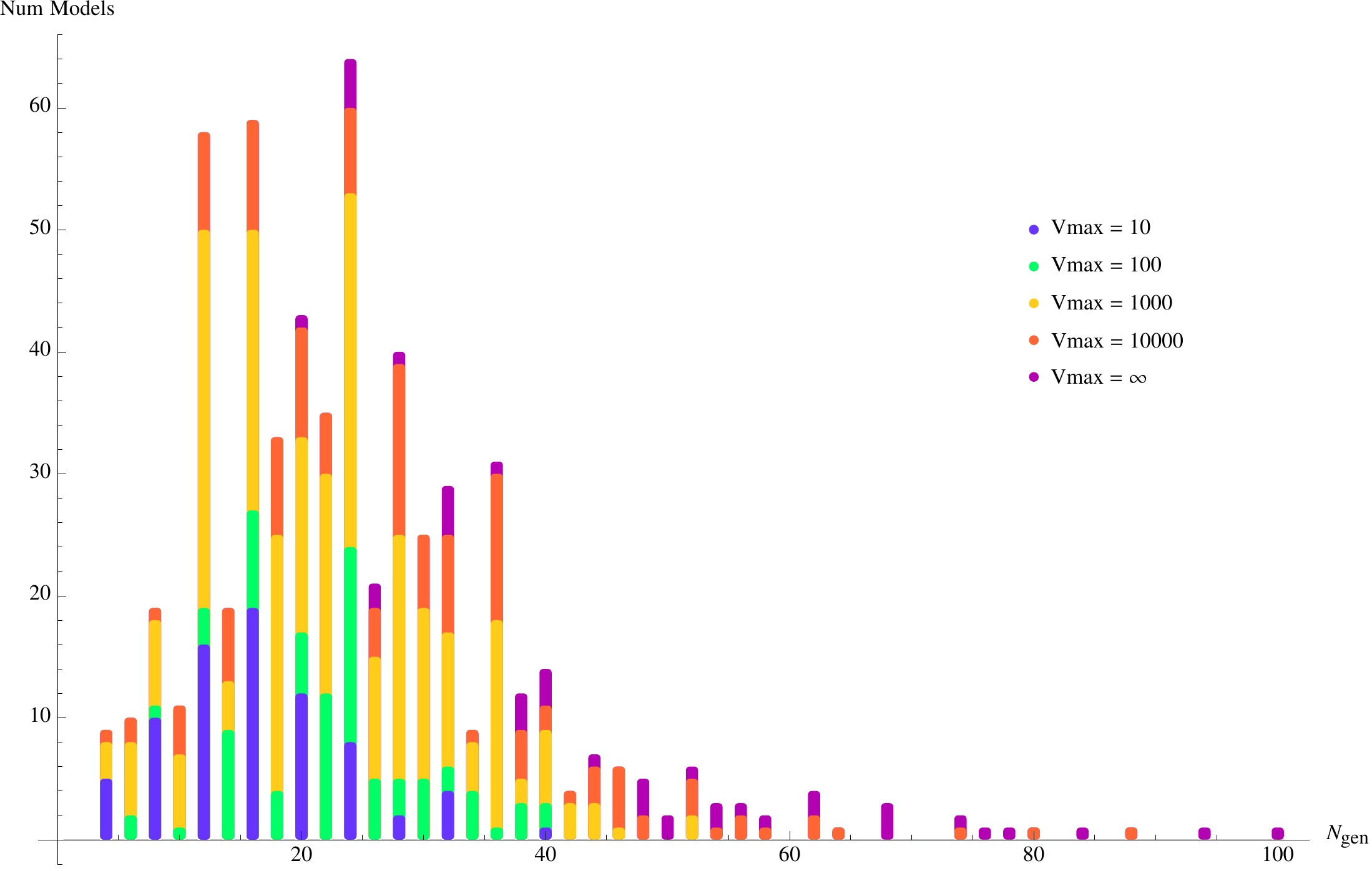}
\capt{16cm}{fig:plot7862}{Plot of the number of $SU(5)$ line bundle models on the tetra-quadric manifold as a function of the number of generations, $N_{\text gen}(X)$. Different colours correspond to different values of $V_{\rm max}$. Note that the manifold does not admit any models with an odd number of generations, due to the fact that all intersection numbers $d_{ijk}$ are divisible by $2$. However, smooth quotients of $X$ do admit models with an odd number of generations, in particular $3$.}
\end{center}
\end{figure}
\vskip 5mm
\begin{figure}[!h]
\begin{center}
\includegraphics[width=12cm]{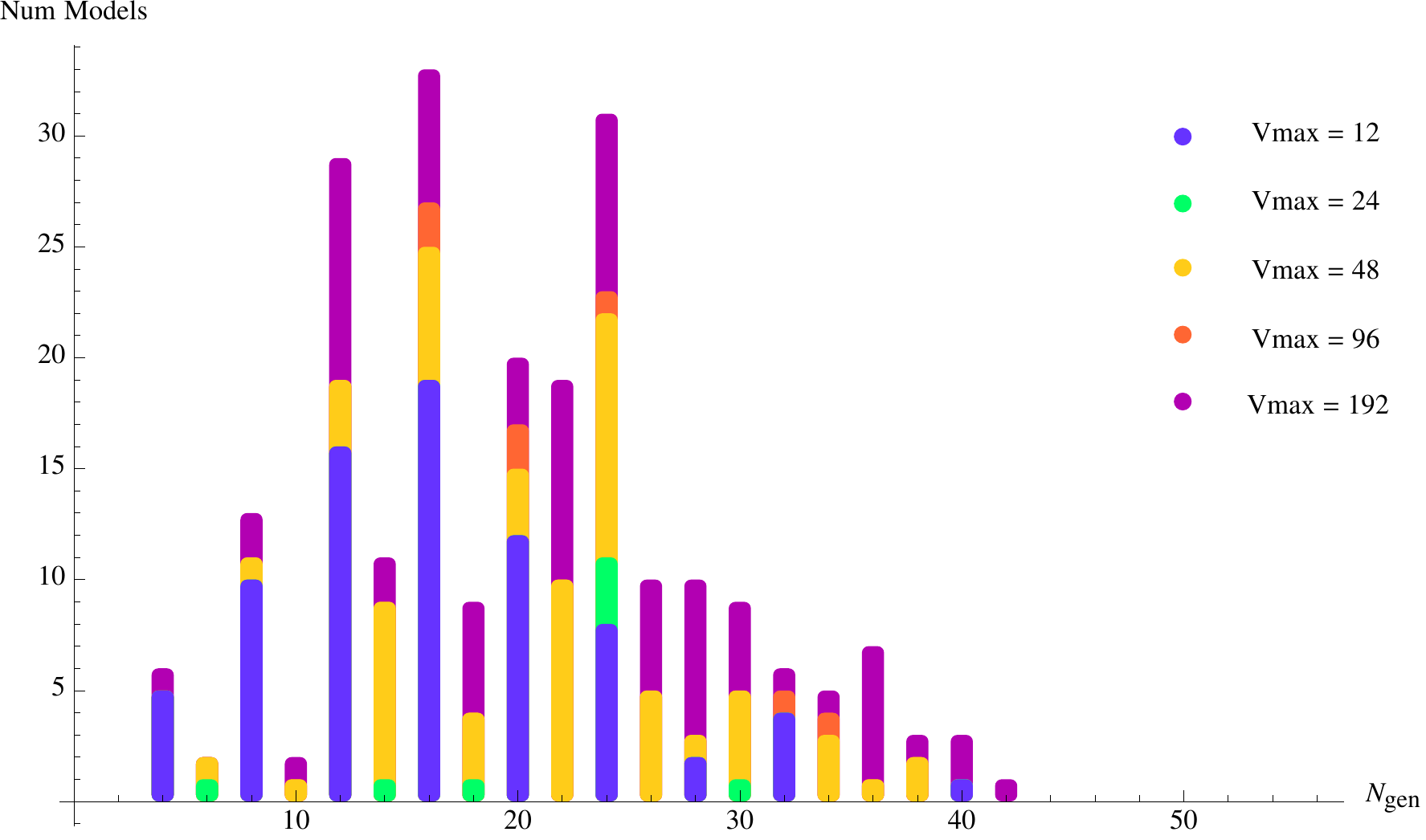}
\capt{16cm}{fig:plot7862v2}{Plot of the number of $SU(5)$ line bundle models on the tetra-quadric manifold as a function of the number of generations, $N_{\text gen}(X)$. Different colours correspond to different values of $V_{\rm max}$.}
\end{center}
\end{figure}
\vskip -5mm\noindent

\subsection{Manifolds with $h^{1,1}(X)=5$}
It is interesting to study the distribution of generation numbers for manifolds with different values for $h^{1,1}(X)$. In this section we present the results for two complete intersection Calabi-Yau manifolds with $h^{1,1}(X)=5$. However, we will not be able to carry out the dependence of the upper bound for the number of generations on the Calabi-Yau volume, due to the technical difficulty of the problem, as explained above. 

\vspace{12pt}
The first manifold is defined by the following configuration matrix:
\begin{equation}\label{eq:X2}
X~=~~
\cicy{\IC\IP^1 \\   \IC\IP^1\\ \IC\IP^1\\ \IC\IP^1\\ \IC\IP^3}
{ ~2&0&0&0 \!\!\!\!\\
  ~0&2&0&0\!\!\!\! & \\
  ~0&0&2&0\!\!\!\! & \\
  ~0&0&0&2\!\!\!\! & \\  
  ~1&1&1&1\!\!\!\! }_{-64}^{5,37}\
\end{equation}
\begin{figure}[!ht]
\begin{center}
\includegraphics[width=15cm]{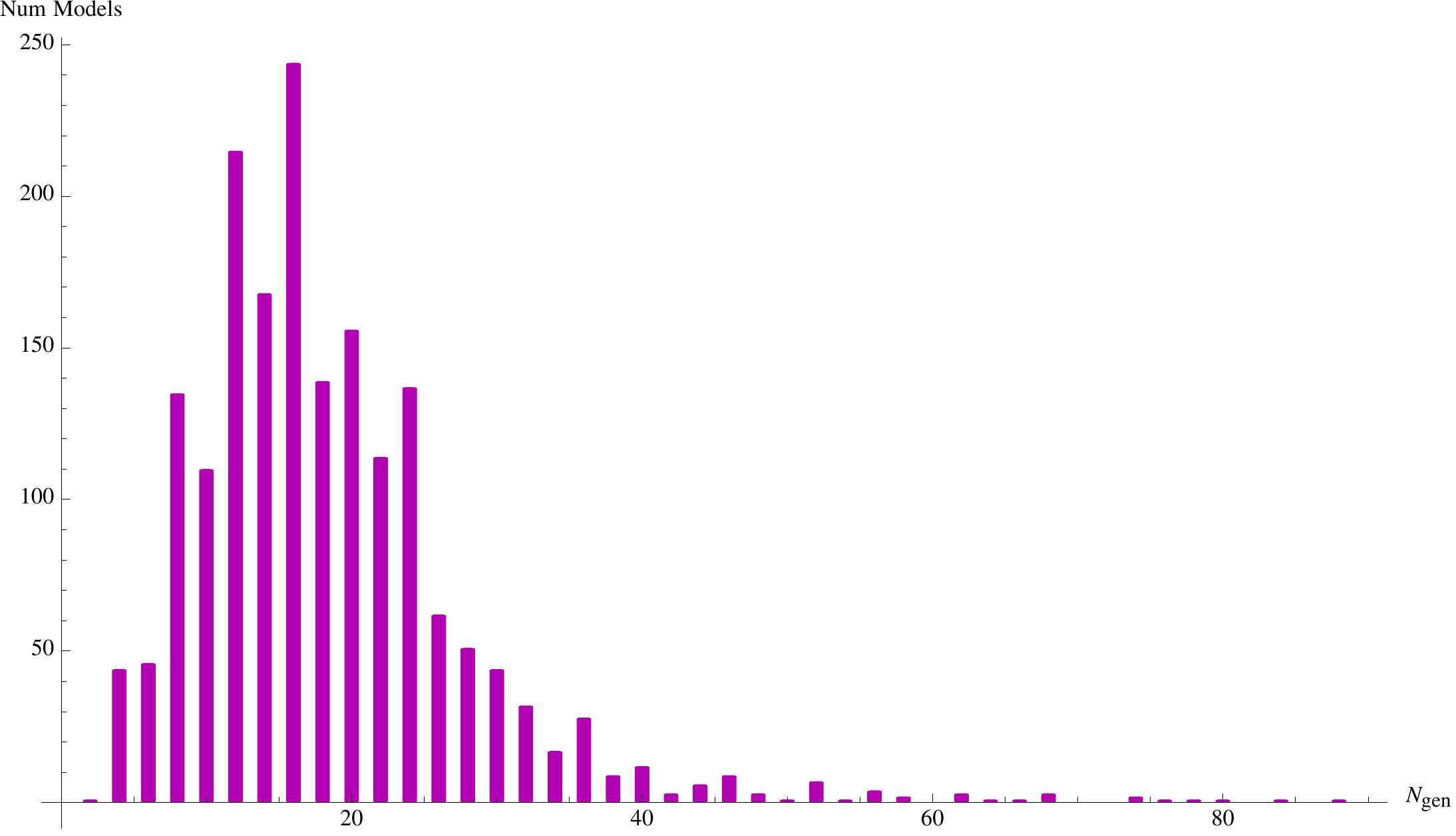}
\capt{15cm}{fig:plot6836}{Plot of the number of $SU(5)$ line bundle models on the manifold~\eqref{eq:X2} as a function of the number of generations, $N_{\text gen}(X)$. The manifold does not admit line bundle models with an odd number of generations, since all its intersection numbers are divisible by 2. }
\end{center}
\end{figure}
\vskip -5mm\noindent
For this manifold the upper bound on the number of generations is $N_{\rm gen}\lesssim 90$, with the distribution peaking in the range $5-25$ generations. Since the manifold admits quotients by finite groups of orders $2,4,8$ and $16$, the distributions for the corresponding quotients peak at considerably lower values, consistent with $N_{\rm gen} = 3$. 

\vspace{12pt}
The second manifold is defined by the following configuration matrix:
\begin{equation}\label{eq:X3}
X~=~~
\cicy{\IC\IP^1 \\   \IC\IP^1\\ \IC\IP^1\\ \IC\IP^4\\ \IC\IP^4}
{ ~1&1&0&0&0&0&0&0 \!\!\!\!\\
  ~0&0&1&1&0&0&0&0\!\!\!\! & \\
  ~0&0&0&0&1&1&0&0\!\!\!\! & \\
  ~1&0&1&0&1&0&1&1\!\!\!\! & \\  
  ~0&1&0&1&0&1&1&1\!\!\!\! }_{-64}^{5,37}\
\end{equation}

The distribution of models as a function of the number of generations is presented in Figure~\ref{fig:plot6724}. The manifold admits a quotient by $\IZ_2$, with Hodge numbers $h^{1,1}(X)=4$ and $h^{2,1}(X)=20$ (see \cite{Candelas:2014} for the computation of the Hodge numbers).
\begin{figure}[!h]
\begin{center}
\includegraphics[width=15cm]{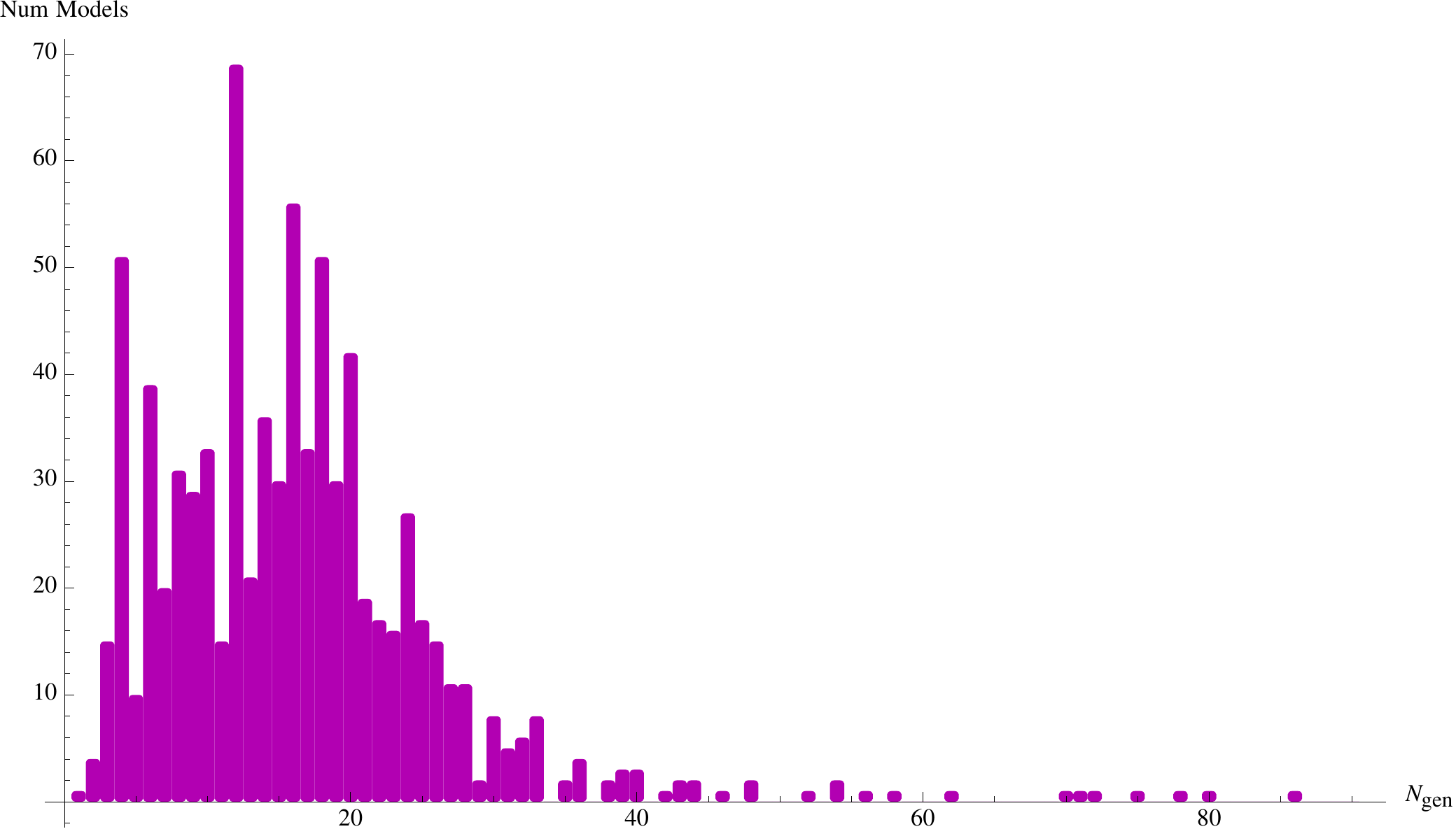}
\capt{15cm}{fig:plot6724}{Plot of the number of $SU(5)$ line bundle models on the manifold~\eqref{eq:X3} as a function of the number of generations, $N_{\text gen}(X)$. }
\end{center}
\end{figure}
\section{Conclusions}\label{sec:conclusion}
In this note, we have provided evidence for the existence of an upper bound on the number of low-energy fermion generations resulting from Calabi-Yau compactifications of the heterotic string with Abelian bundles. Line bundle sums $V$ lead to consistent models provided $c_1(V)=0$ (to allow for an embedding of the structure group into $E_8$), the anomaly condition $c_2(V)\leq c_2(TX)$ is satisfied and the slope zero conditions for all line bundles in $V$ have a common solution in K\"ahler moduli space. We have formulated two versions of the last condition. In the first, mathematical, version all slope zero conditions have to be satisfied in the interior of the K\"ahler cone. In the second, physical, version all slope zero conditions have to be satisfied in the ``physical" region of K\"ahler moduli space. By this we mean a region away from the boundaries of the K\"ahler cone (so that the supergravity approximation is valid) and consistent with a finite Calabi-Yau volume ${\cal V}$, in order to account for the value of the low-energy gauge couplings. 

For the second, physical, version of the slope zero condition, we have shown semi-analytically that the number of consistent line-bundle models on a given Calabi-Yau manifold must be finite. In particular, this means that the number of generations, $N_{\rm gen}$, is finite and is subject to an upper bound which depends on the Calabi-Yau manifold and the maximum value of the volume ${\cal V}$. This suggests a possible correlation between the observed number of generations and the value of the gauge coupling constants: The sizeable value of the GUT coupling constant implies a relatively small Calabi-Yau volume which, in turn, leads to a stringent constraint from the zero-slope conditions and, hence, to a tight upper bound on the number of families. 

Using computer scans, we have explicitly analysed this relation for two Calabi-Yau manifolds: the tetra-quadric Calabi-Yau manifold with $h^{1,1}(X)=4$ and a couple of other complete intersection Calabi-Yau manifolds with $h^{1,1}(X)=5$. In both cases we have found that the number of models is finite even when the weaker, mathematical version of the slope zero conditions is used. In the context of the stronger, physical version of the slope zero conditions, we have also determined the maximal number of generations as a function of the maximal volume. In agreement with general expectations, a small Calabi-Yau volume, ${\cal V}\simeq 12$, as required to account for the physical value of the GUT gauge coupling, implies a stringent bound of $N_{\rm gen}\lesssim 40$ for the tetra-quadric , with peaks of the relevant distributions at significantly lower values. These numbers still have to be divided by the order of freely-acting symmetries to obtain the physical number of generations and are, therefore, easily consistent with three generations.

\section*{Acknowledgements}
AC~is partially supported by the COST Short Term Scientific Mission MP1210-28996 and would like to thank the Theoretical Physics Department at Oxford University for hospitality during part of the preparation of this paper. AL~is partially supported by the EPSRC network grant EP/N007158/1 and by the STFC grant~ST/L000474/1.  CM~would like to thank the Rhodes Scholarships and the Yusuf and Farida Hamied Foundation for support during this work.



\newpage
\bibliography{bibfile}{}
\bibliographystyle{utcaps}

\end{document}